\documentclass[letterpaper]{article} 
\usepackage{aaai2027}  
\usepackage[hyphens]{url}  
\usepackage{graphicx} 
\usepackage{natbib}  
\usepackage{caption} 
\usepackage{textcomp}
\usepackage{booktabs}

\newcommand{\name}{\textbf{IUU+DB}}

\title{IUU+DB: Tracking Illegal, Unreported, and Unregulated Fishing,\\
Seafood Fraud, and Labor Abuse through LLM-Driven Information Extraction}

\author{
    Henry Bodwell\textsuperscript{\rm 1},
    Hong Yang\textsuperscript{\rm 1},
    John C. Simeone\textsuperscript{\rm 2},
    Kelvin Gorospe\textsuperscript{\rm 3},
    Bella Sullivan\textsuperscript{\rm 4},\\
    Lana Huang\textsuperscript{\rm 5},
    Jessica Gephart\textsuperscript{\rm 5},
    Sandy Aylesworth\textsuperscript{\rm 4},
    Molly Masterton\textsuperscript{\rm 4},
    Naren Ramakrishnan\textsuperscript{\rm 1}
}
\affiliations{
    \textsuperscript{\rm 1}Virginia Tech, United States\\
    \textsuperscript{\rm 2}Simeone Consulting, LLC, United States\\
    \textsuperscript{\rm 3}Independent, United States\\
    \textsuperscript{\rm 4}Natural Resources Defense Council, United States\\
    \textsuperscript{\rm 5}University of Washington, United States\\
    henrybod@vt.edu, yanghd@vt.edu, simeoneconsulting@gmail.com, kdgorospe@gmail.com,\\
    isullivan@nrdc.org, lrhuang@uw.edu, gephart@uw.edu, saylesworth@nrdc.org,\\
    mmasterton@nrdc.org, naren@vt.edu
}

\begin{document}

\maketitle

\begin{abstract}
Illegal, unreported, and unregulated fishing (IUU) traditionally refers to fishing activities that violate applicable laws or occur in areas that lack applicable laws. We propose the term IUU+ to capture a broader suite of fisheries sector environmental and associated supply chain trade-related crimes and behaviors.
Although IUU+ activity is widely recognized as a serious threat to marine ecosystems, markets, and livelihoods,
a quantitative understanding of these incidents, e.g., their frequency, geography, species, actors, and patterns in the type of illicit activity, remains difficult to obtain.
We propose IUU+DB, a large language model (LLM)–driven system for building a global incident database of IUU+ activity. The system ingests heterogeneous documents, classifies whether they describe relevant incidents, extracts key data elements such as actors, locations, species, vessels, violations, and enforcement outcomes, and supports deduplication and trend analysis.
Case studies and validation results show that IUU+DB can help organize fragmented evidence, surface geographic and behavioral hotspots, support fisheries-domain specific research in academia and non-government organizations, assist source and species risk assessments for industry, and provide support for policy implementation and targeted enforcement efforts to government agencies.
\end{abstract}


\section{Introduction}
Illegal, unreported and unregulated fishing (IUU) undermines sustainable fisheries management which relies on capping harvests, limiting the use of destructive gears, limiting (or minimizing) harvest of juveniles, and avoiding fishing in sensitive ecosystems, among other regulations. While it is difficult to quantify the full extent of illegal fishing, a 2009 estimate of global losses from IUU fishing amounted to \$10 bn and \$23.5 bn annually, representing between 11 and 26 million metric tons~\cite{agnew_estimating_2009}.

While IUU fishing encompasses a broad set of activities, there is a broader set of associated or analogous activities across the seafood sector, including illegal aquaculture, mislabeling, labor abuses, and trade prohibitions and sanctions evasion. We therefore offer a more inclusive term (IUU+) to capture the broader suite of human, environmental, and associated supply chain-related crimes and behaviors. IUU+ behaviors threaten both people and nature, and undermine key aspects of the UN Sustainable Development Goals~\cite{auld_collective_2023}. The consequences of IUU+ behaviors and their associated crimes are far-reaching: destruction of marine and freshwater ecosystems, loss of human livelihoods and socioeconomic security, safety and human rights of fishing communities, loss of local and national revenues, and unfair competition to legitimate fishermen and businesses~\cite{shaver_casting_2018}. And the impacts of IUU+ behaviors extend beyond the fisheries sector as they have been linked to transnational organized crime, illicit financial trade such as trade-based money laundering, and circumvention of trade policies such as tariffs, sanctions, and import prohibitions~\cite{may-mavrellis_transnational_2017}.  

To deter, detect, and combat IUU+ behaviors and associated trade, governments have responded with a range of regulatory initiatives that span from on-the-water fisheries monitoring, control, and surveillance (MCS) procedures, to port controls (e.g., Port State Measures Agreement, PSMA), to international trade and import control schemes (e.g., U.S. Seafood Import Monitoring Program, and the European Union's IUU Fishing Regulation)~\cite{un_fao_monitoring_2026,un_fao_agreement_2026,noaa_fisheries_simp_2025,european_commission_eu_2026}. In addition, members of the seafood  industry partnered with international organizations to develop a framework for interoperable catch and supply chain data standards through the Global Dialogue on Seafood Traceability (GDST)~\cite{gdst_global_2026}. Many of these programs rely on the collection and transfer of key data elements (KDEs) across supply chain nodes such as vessel identity, harvest location, species, harvest weight, fishing authorization number, and chain-of-custody information~\cite{blaha_guidance_2023,eu_iuu_coalition_import_2025,hosch_advancing_2025}. 

Despite widespread consensus that IUU+ fishing and related crimes occur globally, there
does not exist a comprehensive incident database that systematically catalogs incidents across jurisdictions, source types, and categories of illicit behavior. Some information is reported in the news media while others
are scattered across government announcements of enforcement actions, and still others are in detailed academic publications and NGO reports.
The lack of such a comprehensive incident database makes it difficult to obtain a snapshot of fisheries crime, emerging trends and patterns, and where to invest limited surveillance and enforcement resources.

To address these needs, we developed \name,
a global incident database for illegal fishing, seafood fraud, labor abuse, and related illicit activities. Our goal is to 
leverage advances in LLM-driven information 
extraction technology 
to capture incidents, their attributes, taxonomies of misconduct, and relevant KDEs. Such a resource could help provide near-real-time evidence of IUU+ behaviors and how these behaviors are expressed, differentiated, and connected across species and geographies. Such a system would support consistent analysis and provide actionable evidence for researchers and policymakers. For example, under the U.S.' import control regulation (SIMP), a recent proposal by the overseeing regulatory agency offers the potential to have a two-tier system for seafood import KDE reporting requirements where the agency could periodically review and adjust the species list for each tier based on risk analysis, moving species between tiers as needed~\cite{noaa_fisheries_action_2024}. For such a system to work, information and data related to risks of IUU fishing, seafood fraud, and potential labor abuses would need to be tracked in near-real time through a comprehensive incident
database such as the one described here~\cite{simeone_us_2026}. 

While LLMs have made it easier to rapidly prototype systems such as is presented here,
reliable extraction of incidents remains
challenging. Source documents can be alternatively scientific or colloquial,
considerably lengthy, feature nuanced distinction between IUU+ categories, and contribute a range of potential KDEs. Deduplication (i.e., knowing whether multiple extractions refer to the same incident) is crucial in event reconciliation. 
While there have been efforts to catalog information on illegal wildlife trade broadly, most work to-date has been focused on enumerating the range of scientific taxonomy involved and is often limited to a specific geographic region. Additionally, these datasets do not contain insights on the underlying crime or illicit behavior committed thus making it difficult to use these datasets to assist in building intervention methods~\cite{stringham_dataset_2021,weissgold_us_2024,marshall_mapping_2025}. Database efforts that try to capture multiple dimensions about an incident and the underlying crime are presently manually curated~\cite{traffic_wildlife_2026,belhabib_spyglass_2026}. There is a need for a scalable non-manual way to catalog and classify incidents. 

Our contributions are:
\begin{enumerate}
    \item We present \name, an LLM-driven information extraction framework that automatically discovers, extracts, and classifies structured incident data from heterogeneous sources including news articles, academic papers, NGO documents, and government reports.
    \item Through \name, we are now able to impute global trends in illegal fishing, across more than 140 countries.
    \item Finally, we show through case studies how \name\ can provide actionable insights for regulators, streamlining seafood supply chain risk assessments.
\end{enumerate}
\section{Background and Related Work}

IUU+ fishing and associated trade includes, but is not limited to, activities such as harvesting that violates national or international laws (illegal), fishing that avoids required reporting or monitoring (unreported), fishing that occurs in areas or under conditions where effective management and enforcement do not exist (unregulated), theft or other illicit fish farm operations (illegal aquaculture), species substitution and mislabeling (fraud), forced labor (labor abuses), and tariff evasion (trade barrier and sanctions evasion). 

Tracking and quantifying IUU+ activity is a difficult task~\cite{un_fao_quantifying_2023}. No comprehensive way to track IUU+ behaviors and associated trade presently exists. Prior research on tracking IUU+ is scattered and is either specific to IUU+ behavior-type~\cite{salas_seas_2026, luque_characterization_2019, temple_illegal_2022} or is specific to a particular geography or species~\cite{glaser_foreign_2019,wwf_illegal_2014}. 

Central to how we organize our information classification and extraction is a granular taxonomy of IUU+ behaviors and KDEs. To enumerate IUU+ behaviors, we surveyed the literature and developed a list of 41 behaviors that fall within the seven themes enumerated above. Regarding KDEs, we  conducted a literature review on seafood-specific supply chains to identify KDEs that are either already established or are proposed in the literature to fill identified information gaps. We developed 14 groups of KDEs, which, taken together have over 100 KDE fields (see Appendix \ref{dataDict}).

\paragraph{LLM-Driven Information Extraction}
The transition from pretrained language models such as BERT to LLMs, has enabled an increase in the possible complexity of data extraction as the problems move from strictly extractive to generative~\cite{zhang-etal-2025-survey}. However, these generative IE methods have notable issues when it comes to reliability and reproducibility~\cite{dathe2026usefulexplorationriskyprecision}. To account for this, earlier work has looked at ways to limit hallucination through schema and grounded ontologies~\cite{Hodgson2026, li2025exploring}. Other approaches extract the exact supporting context alongside the generative IE to evaluate grounding against after the extraction~\cite{Spillias2025}, although this can be costly to verify. 


\section{Methodology}
Functionally, we set the minimum information threshold for what qualifies as an incident of IUU+ fishing and associated trade to the common data elements used in crime script analysis: who (actors), what (methods), when (time), and where (location). To ensure broad coverage of IUU+ incidents, we designed our information extraction pipeline around five primary sources:
(i) Government press releases, (ii) Non-Governmental Organizational reports and press releases, (iii) News articles, (iv) Academic Papers, and (v) Industry journals. These five sources span a variety of input lengths and styles, and we also hypothesized that they will have differing foci of IUU+ incidents.

\subsection{Framework Design}
\begin{figure}
    \centering
    \includegraphics[width=0.9\linewidth]{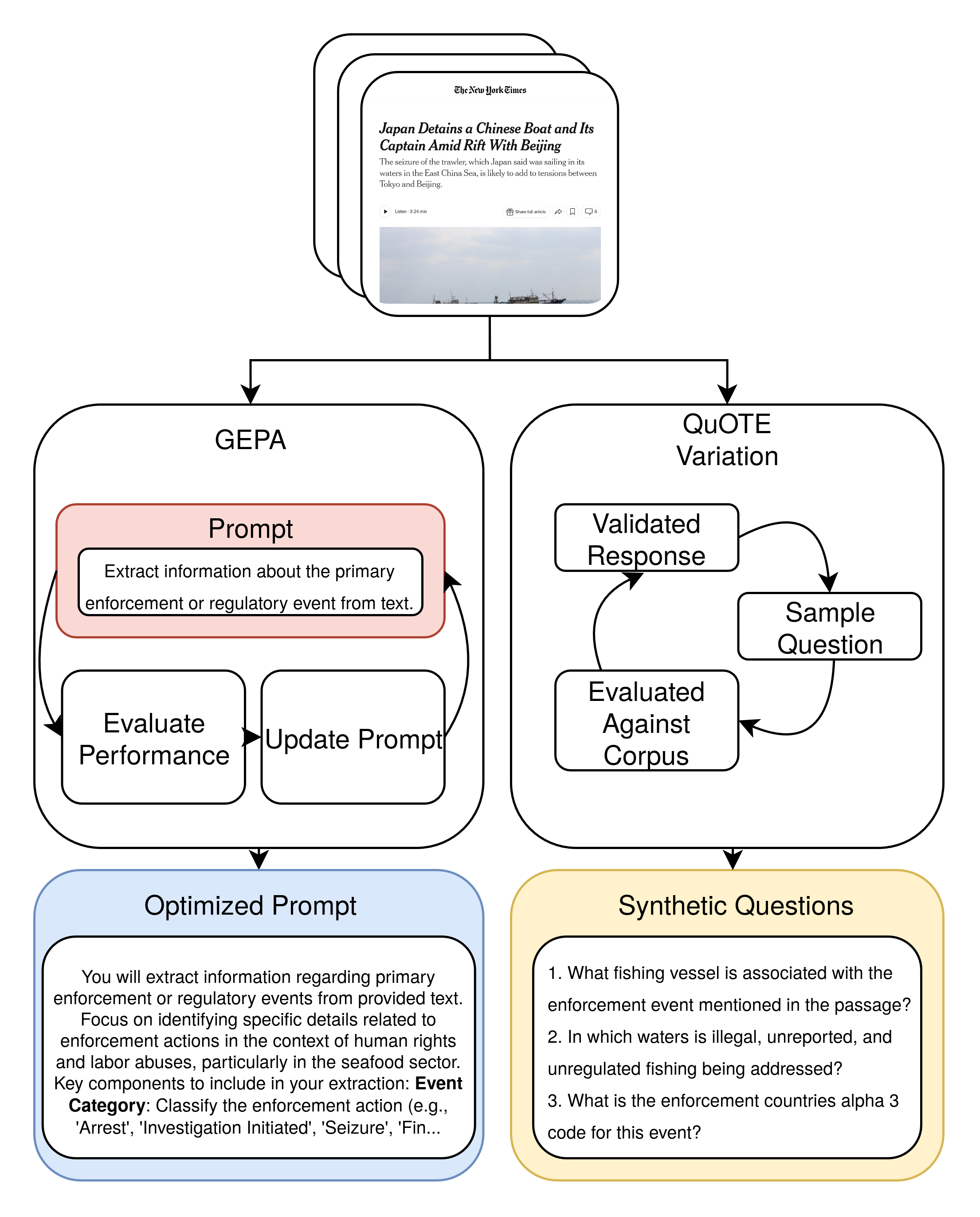}
    \caption{Prompt Optimization Flows. }
    \label{fig:Pipeline}
\end{figure}

\name\ is an end-to-end pipeline for turning unstructured documents into structured incident records. It ingests documents, classifies them by source type, and extracts relevant information into a standard output schema.
The pipeline is guided by an input schema that defines the document categories to recognize and the key data elements (KDEs) to extract. Fig.~\ref{fig:Pipeline} gives an overview of the \name\ pipeline. Based on the KDEs, IUU+ types, and behaviors identified by our team outlined in Appendix   \ref{dataDict}, we classify each source into one of three main source types: (i) sources that describe one or multiple incidents, (ii) sources that are related to IUU+ fishing but do not pertain to an incident, and (iii) sources unrelated to IUU+ fishing. For each incident, the system extracts information from 14 KDE groups covering 100 possible fields, and classifies the incident into 7 possible IUU+ types, covering 41 specific behaviors. 
For sources classified as related to IUU+ but that do not describe a specific incident, the system extracts a set of 3 KDEs.
To manage the large extraction task possible, the pipeline
first aims to ascertain which KDE groups are present in a
document and then extracts fields from only those groups.
This effectively reduces the schema information the LLM needs to hold in context at any given moment.

\subsection{Document Pre-Processing Module}
\textit{Document Pre-Processing} Module is the first step of \name\ and accepts plain text, URLs, and PDFs. For PDFs
we use Pytesseract\footnote{\url{https://pypi.org/project/pytesseract/}} to perform Optical Character Recognition (OCR) to render documents ingested into plain text. For URLs we first attempt to extract text data with Playwright\footnote{\url{https://playwright.dev/python/}}; if that fails,
we use an LLM to extract the core text. Once the text is extracted, the system removes formatting artifacts, checks for duplicate documents using a text hash, and skips documents that have already been processed.
Finally, \name\ splits the cleaned text into smaller chunks, embeds them using Qwen3-VL-Embedding-2B\footnote{\url{https://huggingface.co/Qwen/Qwen3-VL-Embedding-2B}}, and stores them in a vector database for later retrieval. We chose this Qwen model as it supports a token length of up to 32k which is more than enough and supports input that includes a mixture of text and image data, so in the future we can incorporate embedded figures from sources. This step can be parallelized so that large document collections can be processed efficiently.


\subsection{Source Classification Module}
The \textit{Source Classification} Module 
determines what kind of information a document contains.
Given a document and a set of possible source types, the module first identifies passages that appear relevant to IUU+ activity.

Our framework uses few shot prompting to classify the document as one of the following: a source describing an IUU+ incident, a source related to IUU+ fishing but with either more general information or not enough sufficient information provided to be considered a specific incident, or a source unrelated to IUU+ fishing.
Because a single document may describe more than one incident, this module also extracts and stores the passages associated with each incident. This helps the KDE extraction module process each incident separately and reduces confusion in later stages.



\subsection{KDE Extraction Module}
The \textit{KDE Extraction} Module is the workhorse of the
\name{} system. It takes the source document, scope classification, as well as optionally relevant passages and a list of KDE groups. The extractor then makes $n$ runs where $n$ is the number of relevant passages, or $1$ if the classifier did not include them. In each run the extractor first determines the presence of each KDE group given before running the extraction of each KDE fields within the group. If the KDE groups were not given the module defaults to extracting every field. For our implementation we again utilized an LLM with DSPy\footnote{\url{https://dspy.ai/}}. DSPy helps us define a structured schema for extraction and encourages logical consistency across fields. For example, if the system identifies a seafood product, we expect it to also identify the species from which that product is made. From the three source classification, we identify two schema to extract: one for incidents, and another albeit much smaller schema for related to IUU+ fishing which tracks species, countries, and companies mentioned. We do not capture any additional information from unrelated articles. In our experience, we found that some KDEs are relatively rare, and unlikely to occur together within the same incident (e.g. It would be uncommon for there to be a single IUU+ incident as we have described them, to include data about both catch AND aquaculture.) We grouped KDEs therefore by those most likely to co occur with one another and extract them in the modules runs. After extracting the KDEs we then classify the IUU+ behavior exhibited in the incident.

\subsection{Deduplication and Trend Identification}
After extraction, the incidents are aggregated in two ways, first by merging duplicate incidents and secondly by grouping related incidents into trends. The \textit{De-Duplication and Trend Identification} modules first identifies similar incidents using cosign similarity against their sources, we then check whether the duplicate candidates have matching identification information such as vessel name, or id and matching event or enforcement dates. If the two incidents pass those filters and have high cosign similarity, the module merges the extracted KDEs. In the cases of conflicts, the module queries all attached sources for the relevant KDE field in order to get a consensus.

\subsection{Prompt Optimization}
We use two forms of prompt optimization in tandem to automate the iterative task of prompt optimization. The first method is to utilize DSPy and GEPA ~\cite{agrawal2026gepareflectivepromptevolution} to fine-tune the prompt and field descriptions themselves. For our metric we assigned rewards for having correctly extracted fields, and for fields that are extracted in both the gold extraction and the generated extraction the score is calculated by the F1 score of extracted token overlap. The second prompt optimizer we run is a variation on QuOTE \cite{neeser2025quotequestionorientedtextembeddings} where for each KDE we generate a set of questions whose answer is the gold extraction, after which we evaluate whether that question improved retrieval on the corpus and the top 3 questions get appended to the prompt. These two steps are composable and we can run either or both optimizers to improve performance.

\subsection{Datasets}
 We collected data  for this project in two primary ways, directly from  APIs  and from websites via webscrapers. We used the following APIs:  NewsAPI~\cite{NewsAPIai}, SerpAPI~\cite{serpAPI}, and CORE API~\cite{Knoth2023-zi}. And also targeted the following websites via webscrapers: MongaBay~\cite{MongaBay}, US DOJ~\cite{DOJ}, Oceana~\cite{Oceana}, US NOAA Fisheries~\cite{noaaFisheries}, and Undercurrent News~\cite{underCurrentNews}. To create queries for candidate source discovery, we used the previously defined IUU+ types and behaviors
 to build queries that would capture the wide range of behaviors (see Appendix \ref{dataDict}). 
 
 For webscraping, we separated websites into two groups by how specific the sites were to  the fish-seafood sector. For more general sites, e,g., MongaBay and DOJ, we searched for the following key words: "illegal fishing", "IUU fishing", "overfishing",  "fishing violation", "illegal catch", "seafood fraud", "forced labor fishing", "illegal aquaculture", and "illegal seafood sanctions". In all these searches, our aim was to capture any relevant articles capturing the illicit behavior as well as narrow the scope to only fishing and associated trade. Since the second group of sites focus solely on the fish-seafood sector (e.g., nongovernmental organization (NGO) websites, NOAA Fisheries, and Undercurrent News), it was the case that if the article referred to an infraction it was highly likely to be about IUU+ behaviors. As such, we selected the following keywords: "violence", "investigation", "coercion", "arrest", "charge", "indict", "fined", "enforce".

 With the three APIs we can use more complex logic and selected the following query: "IUU or seafood mislabeling or ((Transshipment or sanctions or unfree labor or workplace violations or wage theft) and (ship or seafood or fish) and (arrest or criminal investigation or indictment or search or seizure or fine))". From NewsAPI, instead of using keywords, the query was made up of "concepts" which utilize a semantic searching mechanism allowing us to pull articles from different languages, and related articles that may not use the exact key words.

\subsection{Web Interface}
\begin{figure}
    \centering
    \includegraphics[width=\linewidth]{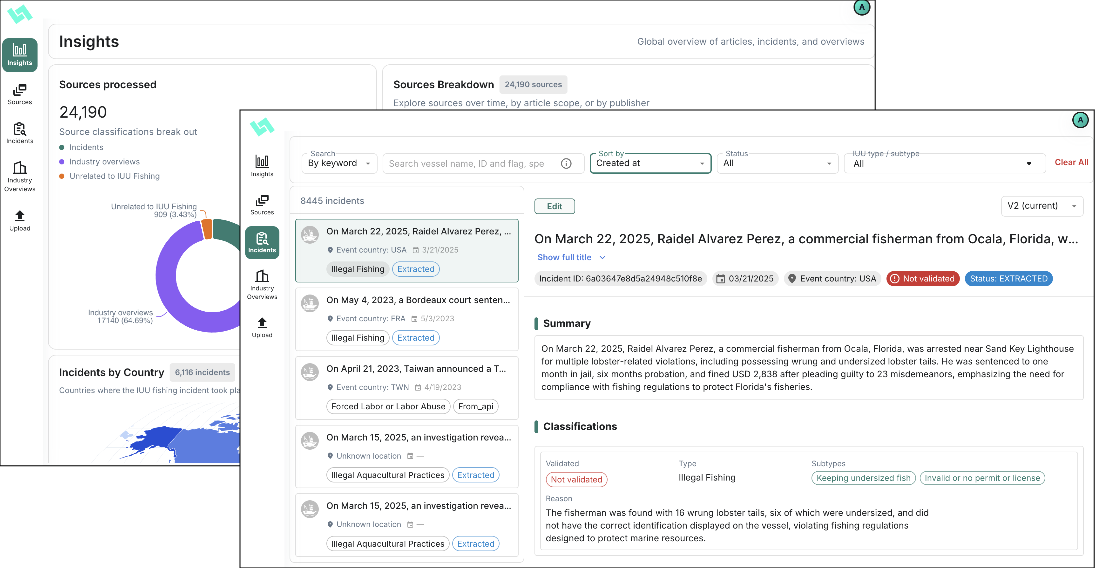}
    \caption{Web UI of the \name\ system.}
    \label{fig:IncidentsPageSC}
\end{figure}

In addition to exposing APIs for general use and further analysis, we developed a web interface to help subject-matter experts analyze, edit, and audit the data. The interface has five main pages. The \textit{Insights} page shows statistics and figures from the database. The \textit{Sources} page allows users to read and explore the documents that have been ingested. The \textit{Incidents} and \textit{Related to IUU+} pages allow users to review the KDEs extracted under each schema. The \textit{Upload} page allows administrative users to add new documents to the database from plain text, URLs, or PDFs. The \textit{Sources}, \textit{Incidents}, and \textit{Related to IUU+} pages also include version history, so administrators can track changes made to each record (see fig \ref{fig:IncidentsPageSC}).


\subsection{Deployment}
Most of the compute is external, so we are able to keep the resource requirements low, \name\ is deployed via docker container on an Ubuntu VM with 100GB of storage, 8GB of ram, and 4 CPU cores. We have 8 admin users with write access. One challenge we faced in deployment was proper logging of site access, as our reverse proxy stripped the IP's of requests when forwarding them to our backend. We fixed this issue when we became aware of it and from what our logs did capture we estimate that we had at least 200 unique site visitors. We had one system outage on for about 3 hours due to the VT servers restarting, to mitigate this risk in the future we are looking to distribute the service across several machines.

The deployment of the \name\ system has been met with great enthusiasm and interest from many fisheries subject matter experts and stakeholder groups, including NGOs, government agencies, academics, fisheries trade analysts, and representatives from multinational fora such as the United Nations Office of Drugs and Crime. At present, fisheries subject matter experts from NGOs and academia are using \name\ to identify risk of specific IUU+ behaviors across fish and seafood supply chains and to promote and provide comment on policies that seek to combat IUU+ fishing and associated trade. In addition, these stakeholders are also using the results from the deployed \name\ to submit funding proposals for further research and development to enable work towards a more regularly updated version, as is described in Section \ref{futureWork} (Future Work). 

\section{Experimental Results}
\name{} ingested articles over 11 years from
2,472 sources across 143 different countries yielding a total of 8,435 incidents.
Our evaluation is focused on addressing the
following questions:
\begin{itemize}
    
    \item Can \name\  be used by SMEs to identify global trends in illegal fishing? (\S \ref{hotspots})
    \item Does the distribution of KDE extractions follow our expectations of IUU+ behavior types? (\S \ref{KDEDist})
    \item Can biclustering on (KDE fields, Behavior types) reveal insights into reporting patterns? (\S \ref{KDEDist})
    \item Can \name\  correctly identify and extract IUU+ incidents and KDEs? (\S \ref{evaluation})
    \item What are the most surprising incidents tracked? (\S \ref{entropy})
    \item Does \name\ demonstrate improved performance vs. newer/more powerful models such as GPT-5.6 Luna? (\S \ref{llmComp})
\end{itemize}

\subsection{Hotspot Analysis}
\label{hotspots}
We find that the data set and \name\ is capturing a wide scope of incidents, from high profile cases~\cite{domPhilips, canCucumber}, to regional conflicts~\cite{LKAIND_1, LKAIND_2, usaSnapper}, to local news~\cite{salmonSpearing, ukFish}. 

\begin{figure}
    \centering
    \includegraphics[width=\linewidth]{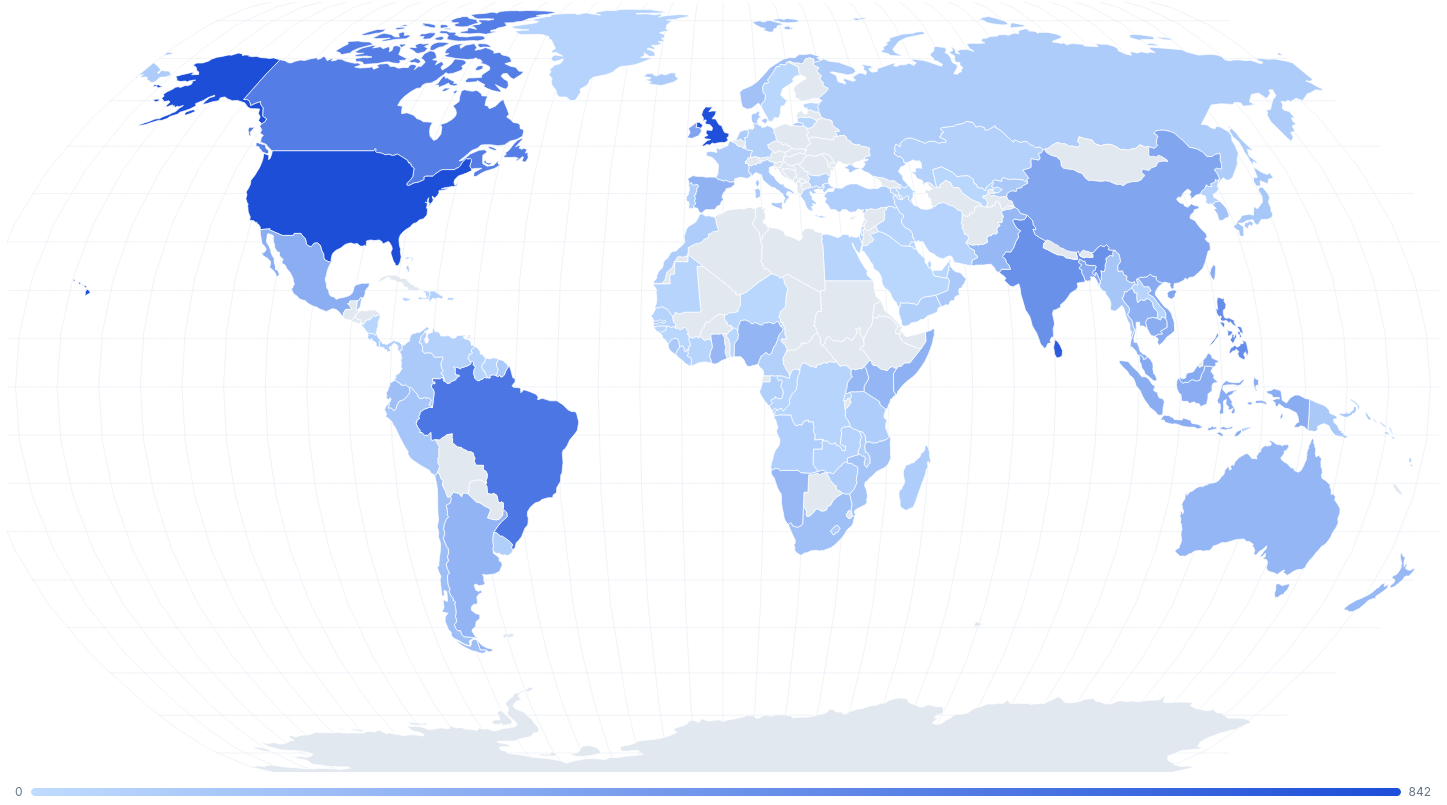}
    \caption{IUU+ Incidents Extracted by Country.}
    \label{fig:Country}
\end{figure}

From Fig.~\ref{fig:Country} we can see that while English speaking nations are highly represented, the dataset has captured events from every continent and nearly every coastal state. The two largest trends we extracted were Illegal Red Snapper harvesting in the Gulf of Mexico and Sri Lankan-Indian fishing rights conflicts.

\begin{table*}[t]
    \small
    \setlength{\tabcolsep}{4pt}
    \centering
    \caption{Incidents extracted via Hotspot Analysis.}
    \label{tab:hotspotSources}
    \begin{tabular}{l|l} \hline
        Country & Incident\\
        \midrule
        Brazil &  Murder of Bruno Pereira \& Dom Philips~\cite{domPhilips}\\
        UK & Illegal Spear Fishing~\cite{salmonSpearing}\\
        Alaska, USA & North Star Fishing Co. Obstruction~\cite{northStar}\\
        Gulf of Mexico, USA & Illegal Red Snapper Harvesting~\cite{usaSnapper} \\
        Nova Scotia & Illegal Eel-Elving Harvesting~\cite{canadaianEel}\\
        British Columbia & 30k Pounds of Tuna Seized~\cite{canadianTuna} \\
        Sri Lanka & Navy Arrests 14 Indian Fishermen~\cite{LKAIND_1}\\
        Philippines & Poaching in Marine Reserve~\cite{phlPoaching} \\ \hline
    \end{tabular}
\end{table*}

\subsection{KDE Prevalence Analysis}
\label{KDEDist}
\begin{figure}[ht]
    \centering
    \includegraphics[width=0.9\linewidth]{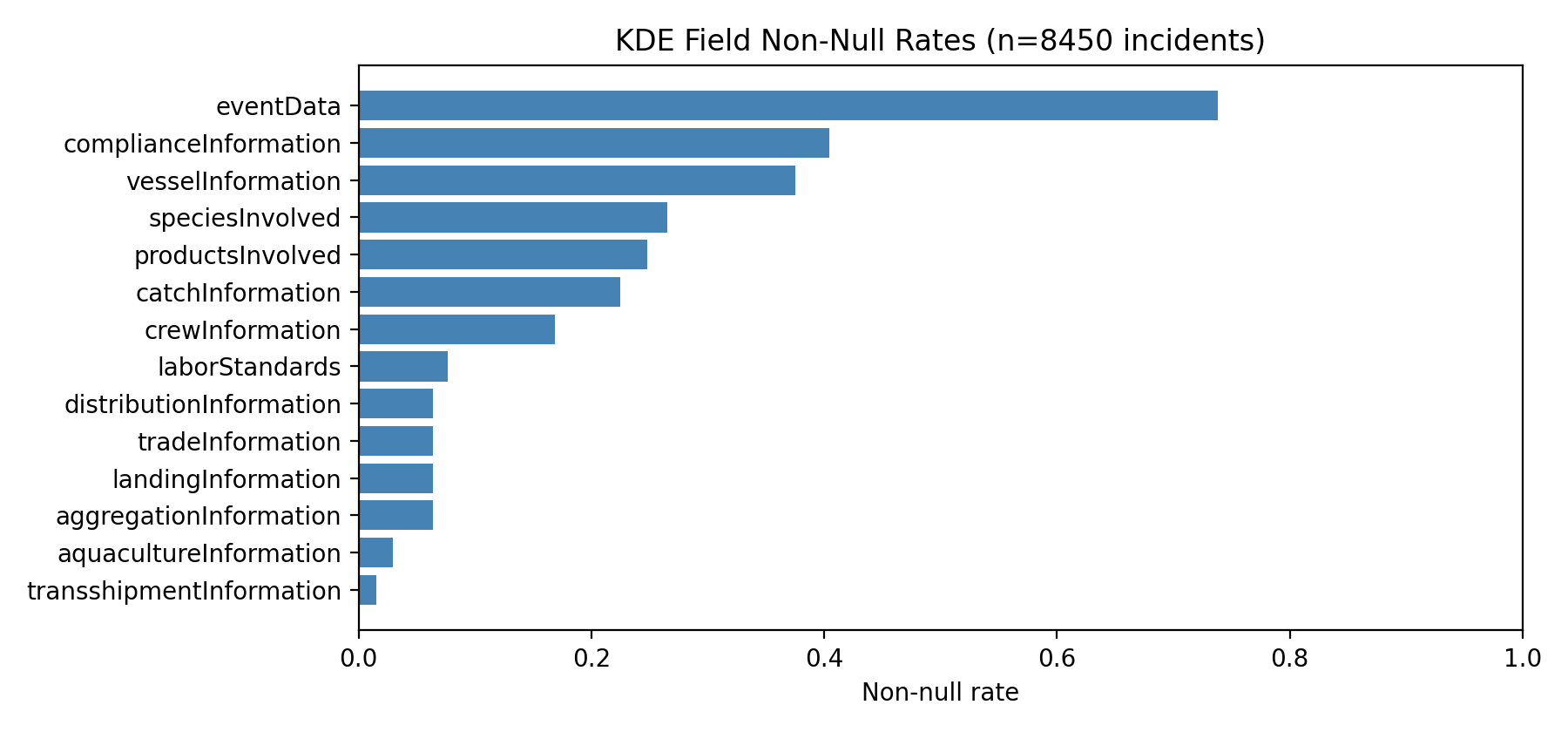}
    \caption{KDE Group Extraction Rate.}
    \label{fig:base_rate}
\end{figure}

\begin{figure}
    \centering
    \includegraphics[width=0.9\linewidth]{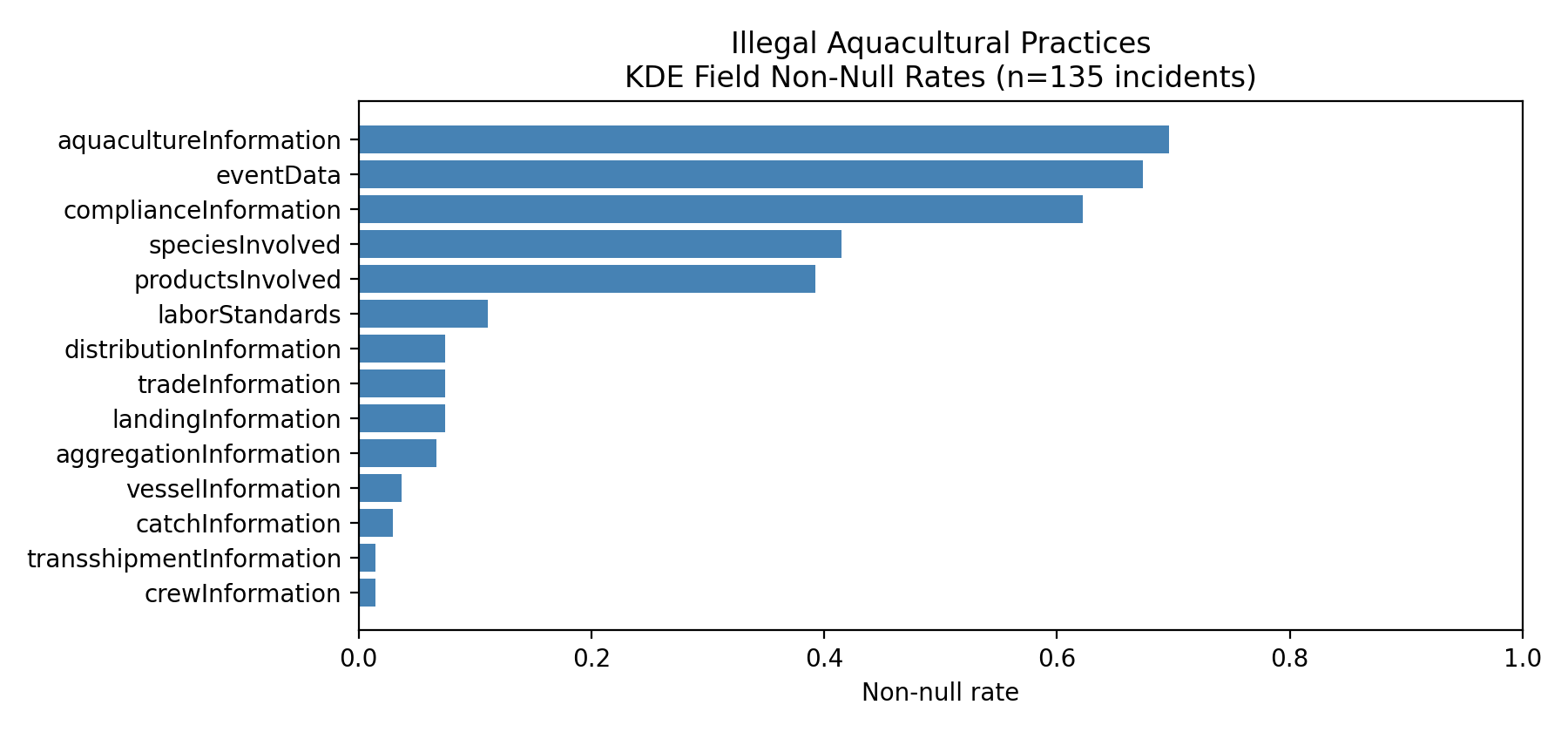}
    \caption{KDE Group Extraction Rate: Aquaculture.}
    \label{fig:aquacultural_rate}
\end{figure}

For a given incident, \name\ extracts roughly 20\% of the fields in the full schema; see Fig.~\ref{fig:base_rate}. This is expected given the large number of fields we are looking for and the kinds of articles that make up most of the database. With a median length of 503 words these articles are often quite short. We next asked whether \name\ finds the KDE groups that we would expect for each IUU+ classification. For this purpose, we analyzed the KDE Presence detection module, which decides whether the system should search for fields within a given KDE group. We then measured KDE group presence across IUU+ types. Figures~\ref{fig:base_rate} and~\ref{fig:aquacultural_rate} show that some KDE groups, such as event and compliance information, appear across many IUU+ types. However, behavior-specific KDEs are much more common in the categories where they are expected. For example, labor standards and crew information increase from 7\% and 16\% across all cases to 81\% and 50\% in labor abuse cases. Similarly, aquaculture information increases from 3\% across all cases to 70\% in aquaculture-related cases.

The second analysis we used biclustering to 
find coordinated occurrences of KDEs and IUU+ sub-behaviors. We use the FP-Max algorithm\footnote{\url{https://rasbt.github.io/mlxtend/user_guide/frequent_patterns/fpmax/}} wherein each transaction is modeled
as a KDE with one or more IUU+ subtypes.
To account for data imbalance, we ran the biclustering in two stages. The first stage was conducted with a minimum support threshold of 0.25 (which yields primarily transactions of IUU+ type of
"Illegal Fishing"), and the second with a minimum support threshold of 0.1 after filtering out transactions from the first step. The maximum set bicluster we found was made up of the following 10 KDEs all from the Event group: event and  enforcement country, event and  enforcement location, and event and  enforcement location category, enforcement category, primary offender, event date, and resolution. This cluster was supported by 45\% of transactions and by 36 unique IUU+ sub-behaviors.

In the second run, there were four clusters each with 10 KDEs, with the most common being the made up of the same 10 KDEs from the first run. The other three were the same as the set above, less event date and with a different KDE instead that is highly relevant to a specific IUU+ behavior or type. Such as, species or product information for Species Fraud and Mislabeling,  or crew information information in cases of Labor Abuse.

\subsection{Maximum Entropy Analysis}
\label{entropy}
We aimed to determine the most surprising illegal fishing incident that occurred within our dataset. For this purpose, we employed a maximum entropy approach. We described each event in terms of an (enforcement country, Illegal fishing sub behavior) tuple. We learn a maximum entropy distribution using iterative proportional fitting (IPF). For each quarter we use incidents from the past year (i.e., past four quarters) to populate the seed matrix which IFP uses to estimate the current quarter. For each cell in the country/sub behavior matrix we then calculated the difference between the estimated and observed, and render it as a z-score. Using $|z| \geq 5$ to define a surprising event, we discover 46 surprising country-behavior events, the top 6 seen in fig \ref{tab:maxent}. Examples of these identified surprising incidents are the USA Q2 2022, and LKA 2024 Q3, the first of which involved two Florida men fishing prior to the season starting and catching undersized fish~\cite{usaLobster}, and the second was a case of falsifying tracking data in Saya de Malha, northeast of Madagascar~\cite{lkaDocuments}.

\begin{table}[t]
    \footnotesize
    \setlength{\tabcolsep}{3pt}
    \centering
    \caption{Top 6 Results from Maximum Entropy Analysis.}
    \label{tab:maxent}
    \begin{tabular}{@{}llp{0.40\columnwidth}r@{}}
        Country & Qtr  & IUU+ Sub Behavior & Z Score\\
        \midrule
        USA & 2022Q2 & Keeping Undersized fish& 26.6\\
        China & 2022Q2 & Falsifying Documents& 25.4\\
        Sri Lanka & 2024Q3 & Falsifying Documents& 17.3\\
        United Kingdom & 2023Q4 & Keeping Undersized fish& 14.2\\
        Ireland & 2022Q2 & Falsifying Documents& 10.4\\
        United Kingdom & 2022Q2 & Prohibited Fishing Gear& 10.3\\
    \end{tabular}
    
\end{table}

\subsection{Evaluation}
\label{evaluation}

For our evaluation set, we had three validators manually label a random selection of 103 source documents, 52 of which were initially classified by \name{} as incidents, 25 were related to IUU+ but not involving an incident, and 26 unrelated to IUU+. The validators read through the document and determined the source scope. If the validator found the document to be an incident, they proceeded to classify the incident from the 7 IUU+ types and 41 sub-behaviors. Finally, the validators extracted the 100 KDEs and entered them into \name{}. We logged any differences and updated the database with the human labeled data.

We first evaluated classification performance on the manually validated test set.  We reported precision, recall, and F1 score for document scope, IUU+ type, and IUU+ sub-behavior. We used the same metrics to evaluate extraction performance at the field level, and then aggregate the results by KDE group. We also reported a mismatch rate. This measured how often the system’s extracted value differs from the validated value when both contained a populated field, for example if the system extracts three species but one of them was spurious that would be considered a mismatch.


\begin{table}[t]
    \small
    \setlength{\tabcolsep}{4pt}
    \centering
    \caption{IUU+ Classification Metrics.}
    \label{tab:Classification}
    \begin{tabular}{lrrr}
        Classification & Precision & Recall & F1 \\
        \hline
        IUU Type (micro)  & 0.82 & 0.87 & 0.84 \\
        IUU Type (macro)  & 0.84 & 0.87 & 0.82 \\
        \hline
        IUU Sub-Behavior (micro) & 0.66 & 0.81 & 0.73 \\
        IUU Sub-Behavior (macro)  & 0.74 & 0.81 & 0.75 \\
    \end{tabular}
    
\end{table}

\begin{table*}[t]
    \small
    \setlength{\tabcolsep}{4pt}
    \centering
    \caption{Source scope classification compared against supervised baselines
    trained}
    \label{tab:sourceClassification}
    \begin{tabular}{l ccc ccc}
    \toprule
     & \multicolumn{3}{c}{Source Scope (macro)} & \multicolumn{3}{c}{Per-class F1} \\
    \cmidrule(lr){2-4}\cmidrule(lr){5-7}
    Model & Precision & Recall & F1 & Incident & Related & Unrelated \\
    \midrule
    TF-IDF + Naive Bayes        & 0.59 & 0.58 & 0.58 & 0.61 & \textbf{0.68} & 0.47 \\
    BERT emb. + logistic reg.   & 0.58 & 0.62 & 0.58 & 0.63 & 0.58 & 0.54 \\
    \name\                      & \textbf{0.70} & \textbf{0.69} & \textbf{0.64} & \textbf{0.67} & 0.59 & \textbf{0.68} \\
    \bottomrule
    \end{tabular}

\end{table*}

\begin{table}[t]
  \centering
  \caption{KDE mismatch rate by group: of the fields populated in both the initial and the final extraction, the share whose values disagree.}
  \label{tab:kde-mismatch-by-group}
  \begin{tabular}{lrrr}
    \toprule
    KDE group & Mismatch  Rate (\%) \\
    \midrule
    Species &  26.67 \\
    Products &  16.67 \\
    Trade &  12.50 \\
    Labor Standards & 11.32 \\
    Catch & 7.41 \\
    Event & 5.63 \\
    Crew & 5.26 \\
    Aquacultural & 0.00 \\
    Compliance & 0.00 \\
    Distribution & 0.00 \\
    Vessel & 0.00 \\
    Transshipment & 0.00\\
    Landing & 0.00 \\
    \midrule
    Overall & 6.82 \\
    \bottomrule
  \end{tabular}
\end{table}

The classification results in Tables ~\ref{tab:Classification} and ~\ref{tab:sourceClassification} show that \name\ performs reasonably well across the main classification tasks. In comparison to the two classification models we trained with the limited training data, \name\ was able to out perform across the board, except in the case against TF-IDF where it outperformed only in the Related to IUU+ but No Incident category. While we outperformed the two very small models a 0.67 in F1 score for incidents leaves a lot of room for improvement. Any gains in stopping false positives on incidents in particular would benefit the system tremendously, both in cutting down the noise to human validators and reducing the load and cost on the system as incidents amount for the majority of LLM calls \name\ makes, up to 7 times the amount for a Related, No Incident article.

\begin{table}[t]
    \footnotesize
    \setlength{\tabcolsep}{3pt}
    \centering
    \caption{Extraction Metrics.}
    \label{tab:Extraction accuracy}
    \begin{tabular}{lccc}
        \toprule
        KDE Group       & Precision & Recall & F1\\
        \midrule
        Compliance      & 1.00 & 1.00 & 1.00\\
        Landing         & 1.00 & 1.00 & 1.00\\
        Vessel          & 0.93 & 0.86 & 0.89\\
        Event           & 0.91 & 0.87 & 0.89\\
        Transshipment   & 0.89 & 0.89 & 0.89\\
        Aquaculture     & 0.75 & 1.00 & 0.86\\
        Labor Standards & 0.82 & 0.85 & 0.84\\
        Crew            & 0.86 & 0.78 & 0.82\\
        Catch           & 0.81 & 0.78 & 0.79\\
        Trade           & 0.78 & 0.64 & 0.70\\
        Products        & 0.79 & 0.60 & 0.68\\
        Distribution    & 1.00 & 0.50 & 0.67\\
        Species         & 0.69 & 0.58 & 0.63\\
        \midrule
        Macro average   & 0.86 & 0.80 & 0.82\\
        \bottomrule
    \end{tabular}
\end{table}

On sources that were classified as an Incident we measured the extraction; results shown in Table.~\ref{tab:Extraction accuracy}. Generally, \name\ did quite well at extracting relevant fields from each source. We found that that about both mismatches and spurious extractions  each made up about 27\% of the total errors in the set, and the remaining 45\% was made up of the extractor missing it. We also measured the rate at which both the human validator and \name\ disagreed with one another or the mismatch rate.  We found that the species and product fields extracted items relevant to the other field, such as a cod filet into the species field, leading to a higher amount of mismatches and spurious extractions in those two KDEs.

From these statistics we can see that while \name\ generally performs well, there is more work needed to reduce the amount of noise, by improving the precision of the scope classifier and reducing the mismatch rate.

\subsection{Ablation Comparison against Baselines}
\label{llmComp}
To evaluate whether \name\ improves performance we compare the results of the test set above against a base line from a newer, more powerful, model GPT 5.6 Luna~\cite{openai2026gpt5.6card}.
\begin{table*}[t]
    \small
    \setlength{\tabcolsep}{4pt}
    \centering
    \caption{Performance comparison between \name, and GPT 5.6 Luna. }
    \label{tab:Comparison}
    \begin{tabular}{l ccc cc cc}
    \toprule
     & Source & \multicolumn{2}{c}{IUU Type} & \multicolumn{2}{c}{IUU Sub-Behavior} \\
    \cmidrule(lr){2-2}\cmidrule(lr){3-4}\cmidrule(lr){5-6}
    Model & F1 & macro F1 & micro F1& macro F1& micro F1\\
    \midrule
    GPT-5.6 Luna  & 0.56 & 0.66 & 0.77 & 0.48 & 0.56 \\
        \name\    &  \textbf{0.65} & \textbf{0.84} & \textbf{0.82} & \textbf{0.73} & \textbf{0.75} \\
    \bottomrule
    \end{tabular}
    
\end{table*}
As can be seen from Table~\ref{tab:Comparison}, \name\ offers a ~10\% improvement on the initial task of classifying the scope, and saw between 5 and 25\% improvement over the newer  Luna model for IUU+ Type and Sub-Behavior, despite \name\ utilizing a smaller, older model, in GPT-4o Mini \cite{openai2024gpt4ocard}.


\section{Ethical Considerations}

\noindent
{\bf Intellectual Property.}
This project aggregates IUU+ content from a range of sources, including some that are behind paywalls or have restrictions on rehosting under their Terms of Service. When paywalled articles were used, we accessed them through legitimate paid subscriptions or purchased access; the system does not scrape or redistribute protected content without authorization. To respect publisher rights and revenue models, \name\ does not show full article text to end users. Instead, users see a summary, source metadata, and a link to the original document. Full text is retained only on the backend for indexing, validation, and audit purposes, and access is limited to a small set of administrative users.

\noindent
{\bf Biased Sources.}
The source documents that are aggregated reflect the perspectives of the organizations and states that produce them. There may be cases where the illegality of an action is disputed between states, e.g., fishing in disputed waters. This project does not aim to adjudicate these disputes nor claim that all cases within are facts, rather we aim to record claimed instances of IUU+ activity for analysis. 

\noindent
{\bf Territorial Ambiguity.}
Some incidents in this dataset occur within territories with contested recognition under international law, such as Somaliland. Assigning an incident to a country necessarily requires choices that may favor one political stance over another. For this project, we match locations to ISO-3166 country codes in order to standardize country names. For events that take place within areas that do not have a ISO code we register any location data available and leave the country field empty. 

\section{Discussion}
\subsection{Limitations}
{\bf Source Equivalence and Factual Reliability}
As previously mentioned, a wide variety of sources are ingested, with a range of credibility. We currently treat all sources as equivalently trustworthy, despite large differences in the evidentiary standards (e.g., a small news organization versus a peer reviewed paper). Users should interpret the incidents as claims of incidents rather than ground truthed fact.  Additionally, our data are mostly from English language sources, and refer broadly to events in English speaking countries.

\noindent
{\bf Data Extraction Quality}
\name\ is limited by the quality of the source material when determining KDEs like the location and species involved. For example, some sources do not specify any species details at all~\cite{ukFish}. Others use generic species group names rather than an exact species distinguisher (e.g., names like `tuna', `salmon', `crab', instead of the specific species names such as `yellowfin tuna', `sockeye salmon', `blue swimming crab'). Meanwhile, other sources  include the full scientific name (e.g., `Thunnus albacares' for yellowfin tuna).

\noindent
{\bf Incident Bias}
The presence of incident reporting should not be the only evaluation criterion used to determine a holistic perspective on the occurrence of IUU+ behaviors. The stories that are covered in the news and through online government agency press releases mean that IUU+ behaviors are being detected and enforcement is taking place. However, in geographies and jurisdictions where enforcement is lacking or non-existent, there will be few incidents to catalog. Nevertheless, there still may be IUU+ behaviors occurring but the lack of enforcement means they are poorly cataloged in our database.

\subsection{Future Work}
\label{futureWork}
Future work will focus on improving extraction accuracy and strengthening source-scope and IUU+ type classification. Better classification will reduce noise in the database and make it easier for subject matter experts to identify useful patterns and insights. We will also use the generated labels, along with the a growing set manually labeled data from our users, to develop and evaluate alternative classification models that can improve \name's performance.

We also plan to expand the system to ingest image and figure data. Some important information may appear only in graphics, tables, maps, or other visual formats, and adding this capability would allow \name\ to capture a broader range of evidence from each source.

Another area of future work is the development of a more user-friendly record validation pipeline that can be opened up to external users. As the database grows, expert validation will remain important for improving record quality, correcting errors, and helping the system learn from reviewed examples.

Finally, we plan to increase the rate at with \name\ ingests to content to a weekly basis. This will include improved extraction pipelines for each source type and broader coverage of web-scraped sources, including additional NGO, government, industry, and media sources from around the world.



\section{Conclusion}
This paper offers \name\ as a  flexible framework for extracting complex schema from variable length sources at scale. \name\ is a tool unique in its scope and potential for near-real time of coverage of IUU+ incidents, giving it the potential to be highly useful to regulators or stakeholder groups. Users can use \name\ to quickly compile available recent information on a country's or fishery's IUU+ incidents,  apply risk-based screening to seafood trade controls, or better understand links between KDEs (e.g., certain fishing practices and labor abuses) and underlying IUU+ incidents. These varied uses for IUU+DB can help improve fishery conservation and social outcomes.

Our results also show that \name\ can surface patterns that are difficult to see from individual reports alone. These include the distribution of incidents across IUU+ behavior types, countries, species, and source categories, as well as co-occurrence patterns among IUU+ types and behaviors. Future analyses can use these structured records to identify emerging hotspots, compare reporting patterns across source types, and better understand how different forms of IUU+ activity are connected. This makes \name\ not only a database, but also a tool for generating evidence that can inform policy, enforcement priorities, and future research.




\appendix
\section{Appendix: IUU+ Behaviors and KDE Data Dictionary}
\label{dataDict}
We define 7 IUU+ types each with a set of related behaviors. They are as follows: 1) Illegal Fishing, constituent behaviors include: exceeding catch quotas, keeping undersized fish, catching unauthorized, protected or prohibited species, the use of banned or prohibited fishing gear, fishing in closed areas or during closed seasons, obscuring vessel identity, falsifying documents, engaging in unauthorized transshipment, engaging in illegal bycatch; 2) Unreported Fishing, constituent behaviors include: unreported or underreported target catch weight or size, unreported or underreported discards, bycatch size or weight, misreported target catch species, misreported non target species, misreported gear, misreported fishing time or location; 3) Unregulated Fishing, constituent behaviors include: dishing without a flag state, fishing under flag state not party to RFMO, fishing in unregulated areas or for unregulated stock; 4) Seafood Fraud or Mislabeling, constituent behaviors include: species mislabeling or fraud, production information fraud; 5) Forced Labor or Labor Abuse, constituent behaviors include: wage and pay violations, abusive living conditions, abusing working conditions, inadequate crew size, physical or sexual violence, workplace intimidation, workers families threatened, isolation, migrants threatened; 6) Circumventing Prohibitions or Sanctions, constituent behaviors include: circumventing sanctions (individuals or corporations), circumventing import prohibitions (countries or products); 7) Aquaculture Specific Activities, constituent behaviors include: unapproved or non-native species, illegal sourcing of broodstock or seed, misrepresentation or unauthorized farm operations, unlicensed, unregistered, or unauthorized farm operations, and stolen aquaculture products.

We identified the following 14 groups of KDEs: Species, Product, Event, Vessel, Crew, Labor, Catch, Compliance, Aquaculture, Transshipment, Aggregation, Trade, Distribution, Landing. Taken together these groups have 100 KDE fields.
\bibliography{base}

\end{document}